\documentclass[journal=jacsat,manuscript=article,8pts]{achemso}

\usepackage{chemformula} 
\usepackage[T1]{fontenc} 

\usepackage{graphicx}
\usepackage{caption}
\usepackage{subcaption}
\usepackage{blindtext}
\usepackage{mathtools}
\usepackage{lipsum}
\usepackage{amssymb}
\usepackage{textgreek}
\usepackage{amsmath}
\usepackage{multirow}
\usepackage{array}
\graphicspath{ {./pics/} }

\author{Dinesh Thapa}
\affiliation[]
{Department of Physics and Astronomy, Mississippi State University, MS, USA}
\alsoaffiliation[]
{Center for Computational Sciences, Mississippi State University, MS, USA}
\altaffiliation{\rm~These authors contributed equally to this work}

\author{Junseong Song}
\affiliation[]{Department of Energy Science, Sungkyunkwan University, Suwon, Republic of Korea}
\alsoaffiliation{Center for Integrated Nanostructure Physics, Institute for Basic Science, Suwon, Republic of Korea}
\altaffiliation{\rm~These authors contributed equally to this work}

\author{Vivek Dixit}
\affiliation{Department of Chemistry, Purdue University, West Lafayette, IN, USA}
\author{Santosh KC}
\affiliation{Chemical and Materials Engineering, San Jose State University, San Jose, California, USA}
\author{Bipin Lamichhane}
\affiliation[]
{Department of Physics and Astronomy, Mississippi State University, MS, USA}
\alsoaffiliation[]
{Center for Computational Sciences, Mississippi State University, MS, USA}
\author{Chandani N. Nandadasa}
\affiliation[]
{Department of Physics and Astronomy, Mississippi State University, MS, USA}
\alsoaffiliation[]
{Center for Computational Sciences, Mississippi State University, MS, USA}
\alsoaffiliation[]
{Department of Physics, Bryn Mawr College, Bryn Mawr, PA, USA}
\author{Kimoon Lee}
\affiliation[]{Department of Physics, Kunsan National University, Gunsan, Republic of Korea}
\author{Sung Wng Kim}
\affiliation[]{Department of Energy Science, Sungkyunkwan University, Suwon, Republic of Korea}
\alsoaffiliation{Center for Integrated Nanostructure Physics, Institute for Basic Science, Suwon, Republic of Korea}
\author{Seong-Gon Kim}
\affiliation[]{Department of Physics and Astronomy, Mississippi State University, MS, USA}
\alsoaffiliation[]
{Center for Computational Sciences, Mississippi State University, MS, USA}

\email{sk162@msstate.edu, kimsungwng@skku.edu}

\title[]
  {First principle investigations of the structural, electronic, and phase stability in 2D layered ZnSb}

\abbreviations{IR,NMR,UV}

\begin{document}

\begin{abstract}
Recently, the two dimensional (2D) materials have become a potential candidates for various technological applications in spintronics and optoelectronics. In the present study, the structural, electronic, and phase stability of 2D layered ZnSb compounds of four different phases viz. wurzite($w$), tetragonal ($t$), hexagonal ($h$), and orthorhombic ($o$) have been tuned using the first principle calculations based on density functional theory (DFT). We invoked the Perdew–Burke–Ernzerhof (PBE) functional and the projected augmented wave (PAW) method during all the calculations. Based on our numerical results, we predicted the novel tetragonal phase as stable phase of ZnSb next to existing orthorhombic structure. We reported the pressure induced phase transition between orthorhombic to tetragonal phase at 12.48 GPa/atom. The projected density of states indicates the strong $p-d$ hybridization between Sb-$5p$ and Zn-$3d$ states confirming the nature of strong covalent bonding between them. The electronic band structures suggest that $t$-ZnSb, $w$-ZnSb, and $h$-ZnSb are metallic in nature whereas $o$-ZnSb is semiconducting with narrow band gap of 0.03 eV using PBE. We predicted the possibility of extracting the two dimensional (2D) monolayer sheet in $t$-ZnSb and $o$-ZnSb according to the exfoliation energy criterion. In addition, the 2D monolayer (ML) of $o$-ZnSb has been predicted to be dynamically stable but that of $t$-ZnSb is not stable as manifested in phonon dispersion bands. Surprisingly, the semiconducting band gap nature of $o$-ZnSb changes from indirect and narrow to direct and sizable while going from 3D bulk to 2D ML structure. Further, we estimated the value of work functions for the surfaces of $t$-ZnSb and $o$-ZnSb as 4.61 eV and 4.04 eV  respectively. Such materials can find the niche applications in next generation electronic devices utilizing 2D hetero-structures.
\end{abstract}


\section{Introduction}
In the quest of searching new exotic 2D materials beyond graphene that has unique physical, electrochemical, and optical properties, the structural and electronic properties of the 2D layered ZnSb of four different phases including wurzite ($w$), tetragonal ($t$), hexagonal ($h$), and orthorhombic ($o$) have been investigated in the present work using the first principle calculations. A 2D layered material is the layered stacking of the lattice structure along a fixed crystallographic direction intervened with a large interlayer separation. The discovery of free standing single layer graphene in 2004 ~\cite{novoselov2005two,geim2010rise} has led to the design and synthesis of the new 2D materials with diverse electro-magnetic, thermo-electric, and topological characteristics~\cite{avouris2010graphene,mak2010atomically,splendiani2010emerging,mak2012control,jones2013optical,zeng2012valley}. The successfully synthesized 2D materials include, but not limited to, hexagonal Boron nitride (white graphene)~\cite{song2010large,liu2013plane,takahashi2017structural}, transition metal dichalcogenides (TMDCs) like $\rm {MoS_{2}}$, and $\rm {WSe_{2}}$~\cite{chhowalla2013chemistry,gupta2015recent,xia2014two}, ZnO~\cite{wu2019controlled}, silicene~\cite{oughaddou2015silicene}, germanene~\cite{davila2016few}, stanene~\cite{lyu2019stanene}, black phosphorus~\cite{wu20182d}, and many others are awaiting to be synthesized. The exciting 2D materials like graphene, silicene, stanene, and TMDCs have a myriad of potential applications in optoelectronics, nanoelectronics, and spintronics~\cite{liu2011graphene,sun2010graphene,bao2011broadband,oughaddou2015silicene,lyu2019stanene}. A new class of 2D material MXenes (metal carbides or nitrides) was discovered which is a promising electrode materials for Li-ion batteries, and  capacitors~\cite{lukatskaya2013cation,luo2015bulk,tang20182d}.

Zinc antimonide (ZnSb), first reported in 1948~\cite{almin1948crystal}, belong to the class of II-V binary zintl phase compound. It has been proved already from the previous studies that the bulk orthorhombic structure $o$-ZnSb is an ubiquitous, non-toxic, and promising semiconducting material with its unprecedented thermoelectric, and strong anisotropic transport properties~\cite{PhysRevB.57.6199}. Moreover, it is known for its relative phase stability and high charge carrier mobility and seebeck coeficient ~\cite{kim1998first, bjerg2011enhanced,semizorov1998anisotropic}. The stoichiometric compound, $o$-ZnSb that belongs to the space group $pbca$ ($n^{o}61$) is stabilized in the form of CdSb type with 8 formula units in which (Zn, Sb) atoms occupy the Wyckoff position (8c, 8c) in the crystal lattice with layered rhomboid rings of $Zn_{2}Sb_{2}$ ~\cite{bostrom2004incommensurably,almin1948crystal,mikhaylushkin2005structure}. It has been identified as an electron poor semiconductor with only seven valence electrons per formula unit~\cite{skipidarov2016thermoelectrics}. The electronic band structure of the compound is characterized by multi-valley bands with narrow band gap. In a view to explore graphene like properties in 2D layered ZnSb; Junseong Song et al. successfully synthesized $sp^{2}$ hybridized 2D-ZnSb via the dimensional manipulation of $sp^{3}$ hybridized 3D-ZnSb (bulk), and the selective etching of alkali metals~\cite{song2019creation}. They reported the structural stability of the 2D-ZnSb in an ambient atmosphere, and demonstrated the structural phase transformation from 3D-ZnSb to 2D-ZnSb. Such type of transformation of crystal structure between two different dimensions is the key factor for the recognition of new material or switching the properties of 2D materials. Being intrigued with this fact, we have tailored the theoretical exploration of the structural and electronic properties of the 2D layered ZnSb of four different phases denoted by $w$-ZnSb, $t$-ZnSb, $h$-ZnSb, and $o$-ZnSb with the possibility of pressure induced phase transition among them.

\section{Computational Method}
First principle calculations based on density functional theory (DFT) were performed using the projector augmented wave (PAW) method within the generalized gradient approximation (GGA) as employed in Vienna Ab-initio Simulation Package (VASP)~\cite{package1996g, perdew1992atoms}. The calculations employed the Perdew-Burcke-Ernzerhoff (PBE) exchange-correlation functional within GGA as in our previous works \cite{a0m4,song2021van,lee2021mixed}. The plane wave energy cut-off of 525 eV was used for all the calculations. All the unit cell structures were fully relaxed including the lattice vectors, and atom positions during the structural optimization using the conjugated gradient method until the total energy was converged numerically to less than $1.0\times10^{-8}$ eV per unit cell and the Hellmann-Feynmn force on each atom was less than $10^{-3}$ eV/Å. The Brillouin zone integrations for the geometrical relaxations are performed using $19\times19\times11$ for wurzite and tetragonal, $19\times19\times9$ for hexagonal, and $16\times16\times16$ for orthorhombic structure. We chose the gamma centered grid for the hexagonal, and Monkhorst-pack grid for the tetragonal and orthorhombic structure respectively to enhance the faster convergence. These grids were increased to $31\times31\times21$ for all the 2D-ZnSb structures and $26\times26\times26$ for 3D-ZnSb respectively for the self-consistent calculations to get the accurate description of charge density and density of states (DOS)~\cite{setyawan2010high}. For the geometrical relaxations of the isolated 2D monolayers of ZnSb, the Brillouin zone sampling was performed with $19\times19\times1$ and $16\times8\times1$ respectively with the vacuum length over 20.0 Å such that the inter-layer interaction can be safely ignored. All the calculations were spin polarized using the tetrahedron method of Blöchl correction with the smearing width of 0.05 eV~\cite{higashiwaki2012gallium}. The diagrams of all the geometrical structures, charge densities, and electron localization function (ELF) were produced using VESTA code~\cite{momma2011vesta}. The plots of the first Brillouin zone of each structure showing the high symmetry kpoints were constructed using the program AFLOW~\cite{curtarolo2012aflow}. 

\section{Results and Discussion} 
We divided our calculated results in six different sub-sections I-VI. In I, we discussed the crystal structure and phase transition. In II and III, we calculated the formation and exfoliation energy of four different phases of ZnSb respectively. In IV, we discussed the electronic properties of the materials including charge density difference, Electron Localization Function (ELF), Bader charge analysis, total and orbital projected DOS, and electronic band structures. In V, we studied the thermodynamic stability of the 2D layered bulk and 2D-monolayer structures via the phonon dispersion curves. Finally, in VI, we calculated the work function value of the single layer slab of $o$-ZnSb and $t$-ZnSb.

\subsection{I. Crystal structures and phase transitions}
\begin{figure*}[h]
	\includegraphics[width=1\linewidth]{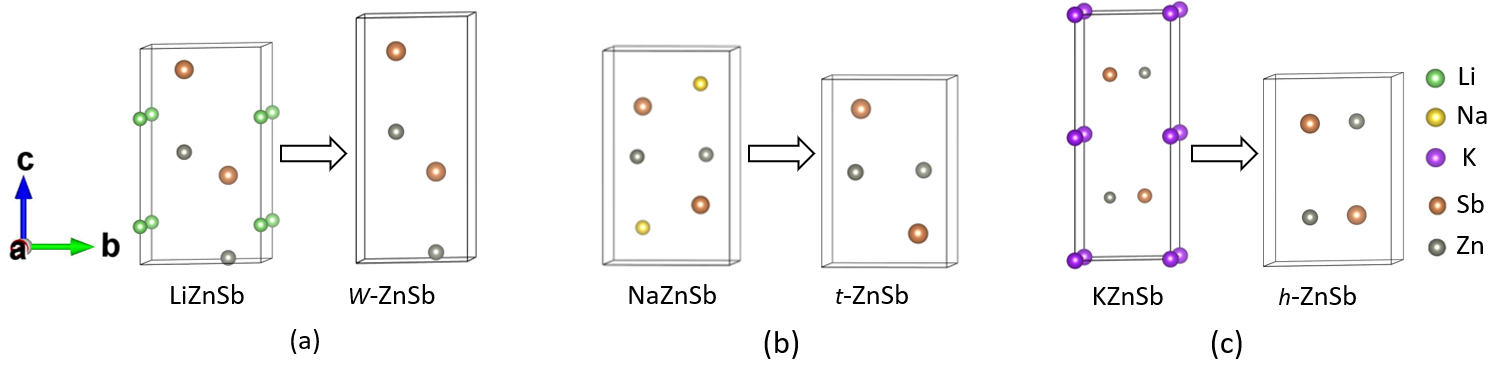}
	\centering
	\caption{Structure of 2D layered ZnSb formed after (a) Li de-intercalation, (b) Na de-intercalation, and (c) K de-intercalation.}\label{Fig:deint}
\end{figure*}

Primarily, the crystal structures used in the calculations were taken from the Inorganic Crystal Structure Database(ICSD)~\cite{ICSD}. We modeled the 2D layered wurzite ($w$-ZnSb), tetragonal ($t$-ZnSb), and hexagonal ($h$-ZnSb) structures respectively by the de-intercalation of the alkali metals (Li, Na, K) from the semiconducting, ternary Zintl phase compounds of the type  $A^{I}B^{II}C^{V}$ where, A = Li, Na, and K; B = Zn, and C = Sb. The compound LiZnSb ($p6_{3}mc$,$n^{o}186$), ICSD No.42064 is stable in wurzite phase with crystallographic Wyckoff position of atoms Li, Zn, Sb at 2a, 2b, and 2c respectively. The compound NaZnSb ($p4/nmm$,$n^{o}129$), ICSD No.12154 is stable in tetragonal phase with Na, Zn, and Sb at Wyckoff position 2c, 2b, and 2c respectively. Similarly, the compound KZnSb ($p6_{3}/mmc$,$n^{o}194$), ICSD No.12161 is also stable in hexagonal phase at 0 K with the atoms K, Zn and Sb at Wyckoff position  2a, 2d, and 2c respectively. LiZnSb and KZnSb shows direct semiconducting behavior in all the computational methods including GGA~\cite{guendouz2017electronic,madsen2006automated,parveen2018topological}; however, the calculated band structure for NaZnSb depends on the method used. It shows metallic behavior within local density approximation (LDA), and generalized gradient approximation (GGA)~\cite{charifi2014phase,jaiganesh2008electronic} but it is predicted to be a semiconductor with a narrow, direct band gap within full potential linear augmented plane wave (FPLAPW) method with the modified Becke–Johnson potential(MBJ), and Cambridge Serial Total Energy Package (CASTEP) program within GGA~\cite{reshak2015thermoelectric,gu2015first}. In this paper, we first carried out the full geometrical relaxation of the compounds AZnSb (A = Li, Na, and K) composed of alternating layers of $A^{+}$ cations and polyanionic structure $(ZnSb)^{-}$ with the layers extending in the $ab$-plane. Our calculated lattice parameters for AZnSb are in close agreement with that of the experimental and previously calculated results  and are shown in Table.~\ref{Table:parent-latt}. Subsequently, we de-intercalated alkali metals from the respective crystallographic sites of AZnSb to form 2D layered structures of $w$-ZnSb (Li de-intercalated), $t$-ZnSb (Na de-intercalated), and $h$-ZnSb (K de-intercalated) as shown in Fig.~\ref{Fig:deint}. Thus obtained ZnSb structures after the de-alkalination will retain the same geometrical structure and Hermann-Mauguin notation as that of their parent compounds with Wyckoff positions for (Zn, Sb) as (2b, 2b) in $w$-ZnSb, (2b, 2c) in $t$-ZnSb, and (2d, 2c) in $h$-ZnSb. The calculated atomic positions of 2D layered ZnSb structures are shown in Table.~\ref{Table:atom-pos}. 

\begin{table}
	\centering
	\caption{The optimized lattice constants ($a$, $c$), axial ratio ($c/a$), equilibrium volume ($V_{o}$), and the value of band gap ($E_{g}$)  for the parent compounds AZnSb (A = Li, Na, K). The experimental values retrieved from the references~\cite{toberer2009thermoelectric,Savelsberg1978on} are given in a bracket in parentheses.\label{Table:parent-latt}}
		\begin{tabular}{l*{6}{c}}
			\hline	
			AZnSb              & $a$ = $b$(\AA)  &  $c$(\AA) & $c/a$ & $V({\mbox{\normalfont\AA}}^{3})$ & $E_{g}$ (eV) \\
			\hline	
			LiZnSb & 4.462(4.428) & 7.240(7.154) & 1.623 & 124.86 & 0.40 \\
			NaZnSb     & 4.466(4.440) & 7.485(7.490) & 1.676 & 149.32 & 0.00 \\
			KZnSb & 4.579(4.540) & 10.731(10.500) & 2.343 & 194.94 & 0.41 \\	
			\hline		
		\end{tabular}
\end{table}

\begin{table}
	\centering
	\caption{Calculated atomic positions of 2D-ZnSb.\label{Table:atom-pos}}
		\begin{tabular}{l*{3}{c}}
		\hline
			Compound    & atom & atomic coordinates  \\		           
			\hline
			\multirow{2}{4em}{$w$-ZnSb} & Zn   & (0.333,0.667,0.028), (0.667,0.333,0.528)  \\ 
			& Sb & (0.333,0.667,0.357), (0.667,0.333,0.857) \\
			\hline
			\multirow{2}{4em}{$t$-ZnSb} &  Zn  & (0.750,0.250,0.500), (0.250,0.750,0.500) \\ 
			& Sb  & (0.250,0.250,0.826), (0.750,0.750,0.174) \\
			\hline
			\multirow{2}{4em}{$h$-ZnSb} & Zn &(0.333,0.667,0.750), (0.667,0.333,0.250)  \\ 
			& Sb & (0.333,0.667,0.250), (0.667,0.333,0.750) \\	
			\hline
		\end{tabular}
\end{table}

\begin{table*}
	\centering
	\caption{The optimized lattice constant ($a$), axial ratio ($b/a$, and $c/a$), equilibrium volume ($V_{o}$), buckling height ($h$), interlayer distance ($d$), bulk modulus ($B_{o}$), pressure derivative ($B_{o}^{'}$), and the value of band gap ($E_{g}$) for 2D-ZnSb, and 3D-ZnSb. The experimental lattice parameters and band gap value (at 300 K) for $o$-ZnSb retrieved from the reference~\cite{arushanov1986crystal} are given in a bracket in parentheses.\label{Table:opt-latt}}
		\begin{tabular}{l*{9}{c}}
		\hline
			Compound              & $a$(\AA)  & $b/a$ & $c/a$ &   $V_{o}({\mbox{\normalfont\AA}}^{3})$ & $h$(\AA) & $d$(\AA) & $B_{o}(GPa)$     & $B_{o}'$ & $E_{g}$ \\
			\hline	
			$w$-ZnSb & 4.004 & 1.000 & 1.982 & 110.19 & 1.36 & 2.61 & 38.50   & 4.41 & 0.00 \\
			\hline
			$t$-ZnSb     & 4.004 & 1.000 & 1.484 & 95.28 &  1.93 & 3.87 & 34.09  & 10.41 &0.00 \\
			\hline
			$h$-ZnSb & 4.576 & 1.000 & 1.331 & 110.45 & 0.00  & 3.05    & 44.44 & 4.92 & 0.00 \\
			\hline
			$o$-ZnSb     & 6.284 & 1.244 & 1.309 & 404.17& - & 2.02 & 47.25 & 5.01 & 0.03 \\
                                    & (6.218) & (1.245) & (1.305) & (390.60) & &  &  &  & (0.50) \\
                                    \hline
		\end{tabular}
\end{table*}

\begin{table}
	\centering
	\caption{The first, second and third nearest neighbor distances ($d_{1}$, $d_{2}$, and $d_{3}$) between two atoms, where $\times$n represents the multiplicity of the bond length. The experimental bond lengths of Zn-Zn, Zn-Sb, and Sb-Sb for $o$-ZnSb retrieved from the reference~\cite{arushanov1986crystal} are given in a bracket in parentheses.\label{Table:bond-length}}  
	\begin{tabular}{l*{5}{c}} 
	\hline
		compound & Atoms & $d_{1}$(\AA)$\times$n & $d_{2}$(\AA)$\times$n & $d_{3}$(\AA)$\times$n \\
		\hline
		\multirow{3}{4em}{$w$-ZnSb} & Zn-Zn  & 4.00$\times$6 & 4.59$\times$6 & 6.09$\times$6  & \\ 
		& Zn-Sb & 2.61$\times$1 & 2.68$\times$3 & 4.78$\times$6  \\ 
		& Sb-Sb & 4.00$\times$6 & 4.59$\times$6 & 6.09$\times$6   &\\
		\hline
		\multirow{3}{4em}{$t$-ZnSb} & Zn-Zn  & 2.83$\times$4 & 4.00$\times$4 & 5.66$\times$4    & \\ 
		& Zn-Sb & 2.78$\times$4 & 4.48$\times$4 & 4.88$\times$8  \\ 
		& Sb-Sb & 3.51$\times$4 & 4.00$\times$4 & 4.80$\times$4   &\\ 
		\hline
		\multirow{3}{4em}{$h$-ZnSb} & Zn-Zn  & 4.03$\times$6 & 4.58$\times$6 & 6.09$\times$2    & \\ 
		& Zn-Sb & 2.64$\times$3 & 3.04$\times$2 & 5.28$\times$3  \\ 
		& Sb-Sb & 4.03$\times$6 & 4.58$\times$6 & 6.09$\times$2   &\\ 
		\hline 
		\multirow{3}{4em}{$o$-ZnSb} & \multirow{2}{2.8em}{Zn-Zn}  & 2.72$\times$1 & 3.74$\times$2 & 4.39$\times$2    & \\
		 &   & (2.59)    &   &    & \\ [1ex]
		&\multirow{2}{2.8em}{Zn-Sb} & 2.69$\times$1 & 2.70$\times$1 & 2.79$\times$1  \\ 
		&   & (2.67) & (2.70) & (2.75) &\\ [1ex]
		&\multirow{2}{2.8em}{Sb-Sb} & 2.84$\times$1 & 3.90$\times$2 & 4.14$\times$2   &\\
		&     & (2.82)  &     &  \\
		\hline
	\end{tabular}
\end{table}

Here, we reported the structural, electronic, and phase transitions in novel 2D layered ZnSb of wurzite ($w$), tetragonal ($t$), and hexagonal ($h$) phases. We also included the structural and electronic properties of the existing orthorhombic ZnSb in order to compare the dimensionality effects. The layered geometrical structures of ZnSb are represented in Fig.2-5 respectively. The unit cell of all the 2D structures comprises of double formula units with 2 Zn atoms and 2 Sb atoms forming a layered structure of $ABAB...$ stacking of ZnSb along the $c$-axis. In $w$-ZnSb structure, 2 Zn and 2 Sb atoms form the double layer in the unit cell where 3 Sb atoms and one Zn atom of the same layer forms tetrahedral coordination with remaining Sb atom of the nearest layer with calculated Sb-Zn-Sb bond angle of $120.48^{o}$ and $96.55^{o}$ as shown in Fig.~\ref{Fig:bond-length}(a). Each layer in $w$-ZnSb is made up of peculiar puckered honeycomb structure with overturned $AB$ stacking of ZnSb. The layers of ZnSb is so structured that Zn atom buckles upward with respect to Sb atom in one layer; whereas, Sb atom buckles downward with respect to Zn atom in next neighboring layer with reference to the crystallographic plane (010), and this pattern repeats alternately in the crystal structure. In $t$-ZnSb, 2 Zn atoms and 2 Sb atoms forms a single layer in the unit cell with a vertex and edge sharing bi-tetrahedral coordination with a Zn atom surrounded by nearest 4 Sb atoms. The two tetrahedra forms a common parallelogram PQRS. The calculated tetrahedral Sb-Zn-Sb bond angle in $t$-ZnSb are found to be $118.88^{o}$ and $91.96^{o}$ as shown in Fig.~\ref{Fig:bond-length}(b). Also, the 2 Sb atoms from each sublayer buckled upward and downward with respect to the middle Zn-plane. The $h$-ZnSb structure is similar to $w$-ZnSb structure except the Zn and Sb atoms align in the trigonal planar structure to form honeycomb layers with the calculated Sb-Zn-Sb bond angle of $120^{o}$ as shown in Fig.~\ref{Fig:bond-length}(c). Similarly, the 2D layered structure in $o$-ZnSb is different from other phases. The layered stacking is made up of quasi-2D layer of rhomboid (parallelogram) rings of $Zn_{2}Sb_{2}$ represented by ABCD in Fig.\ref{Fig:bond-length}(d). The layered structure exhibits the distorted bi-tetrahedra with peculiar five fold coordination of each atom. Here, the two tetrahedra shared the vertices and edges of the common rhomboid ring. The angles between each Sb-Zn-Sb are found to be $109.01^{o}$ which are close to regular tetrahedral angle of $109.5^{o}$. Further, the rhomboids are connected in chains along the $a$-axis through the formation of Sb-Sb dimers to satisfy the octet configuration in which the two layers of rhomboids along the $c$-direction are related by gliding operation. In general, all the crystal structures of ZnSb appear to be composed of tetrahedral units with five fold coordination by one similar and four dissimilar neighboring atoms. 

After the structural optimization of the de-intercalated structures, we observed the significant relative change in $c$ than in $a$ lattice parameters. The axial ratio ($c/a$) increases by $35 \%$ in $w$-ZnSb, decreases by $19 \%$ in $t$-ZnSb, and decreases by $43 \%$ in $h$-ZnSb with corresponding compression in size of all the bulk structures of 2D layered ZnSb with respect to their parent compounds using DFT-GGA. The deviation imposed on the axial ratio $c/a$ in the de-intercalated ZnSb structures is also attributed to the anisotropy in the length of tetrahedrally directed Zn-Sb bond. We reported the optimized lattice parameters ($a$), axial ratio ($b/a$, and $c/a$), equilibrium volume ($V_{o}$), buckling height ($h$), inter-layer distance ($d$), Bulk modulus ($B_o$) and pressure derivative ($B_o^{'}$) for the present structures and shown in Table.~\ref{Table:opt-latt}. It is to be noted that the interlayer distance in 3D-ZnSb refers to the distance between two rhomboid rings along vertical $c$-axis. Also, the calculated value of the first, second, and third nearest neighbor distances $d_{1}$ (\AA), $d_{2}$ (\AA), and $d_{3}$ (\AA) between the species Zn-Sb, Zn-Zn, and Sb-Sb respectively are annotated in Table.\ref{Table:bond-length}.
\begin{figure*}[h]
	\includegraphics[width=1\linewidth]{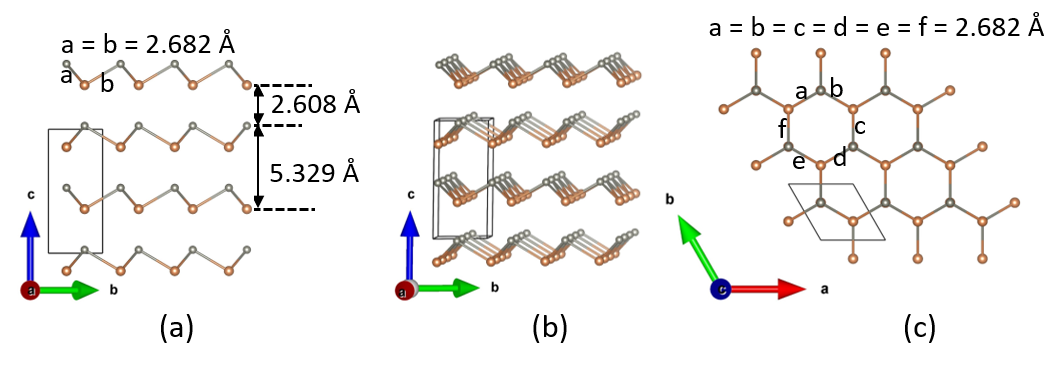}
	\centering
	\caption{Layered 2D structure of Li de-intercalated ZnSb($w$-ZnSb) along plane (a) (100), (b) (100) slightly rotated invariant to $c$ axis, and (c) (001).}\label{w-znsb}
\end{figure*}

\begin{figure*}[h]
	\includegraphics[width=1\linewidth]{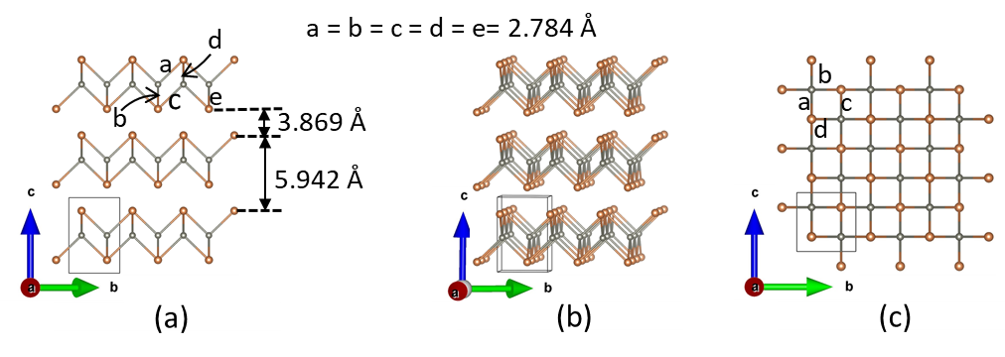}
	\centering
	\caption{Layered 2D structure of Na de-intercalated ZnSb($t$-ZnSb) along plane (a) (100), (b) (100) slightly rotated invariant to $c$ axis, and (c) (001).}\label{t-znsb}
\end{figure*}

\begin{figure*}[h]
	\includegraphics[width=1\linewidth]{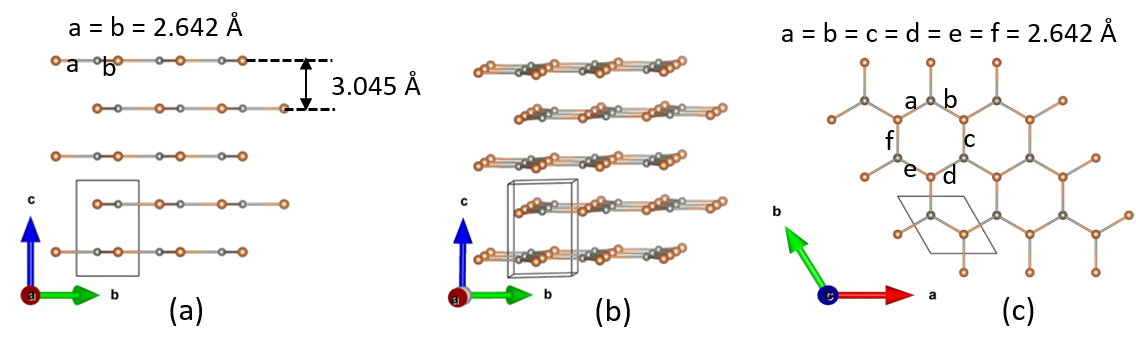}
	\centering
	\caption{Layered 2D structure of K de-intercalated ZnSb($h$-ZnSb) along plane (a) (100), (b) (100) slightly rotated invariant to $c$ axis, and (c) (001).}\label{Fig:h-znsb}
\end{figure*}

\begin{figure*}[h]
	\includegraphics[width=1\linewidth]{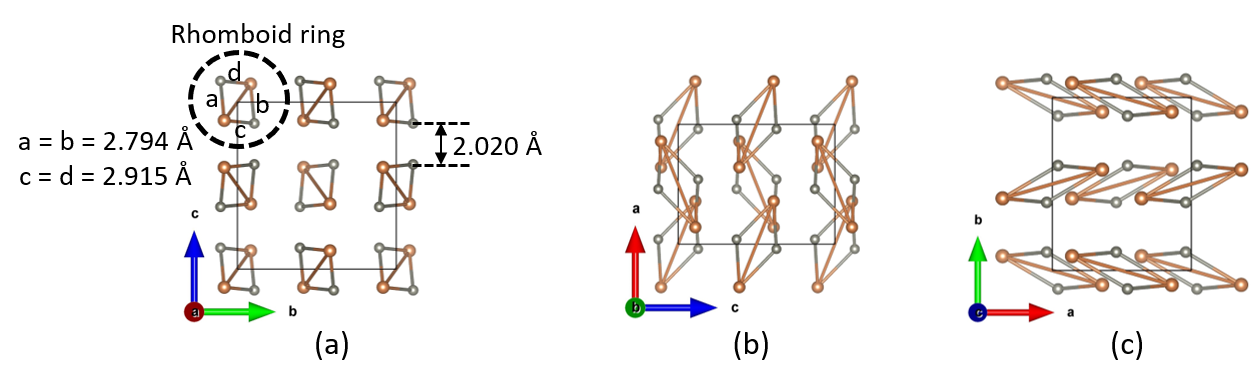}
	\centering
	\caption{Bulk 3D structure of orthorhombic ZnSb ($o$-ZnSb) along plane (a)(100), (b)(010), and (c)(001).}\label{Fig:3D-bulk}
\end{figure*}

\begin{figure*}[h]
\includegraphics[width=1\linewidth]{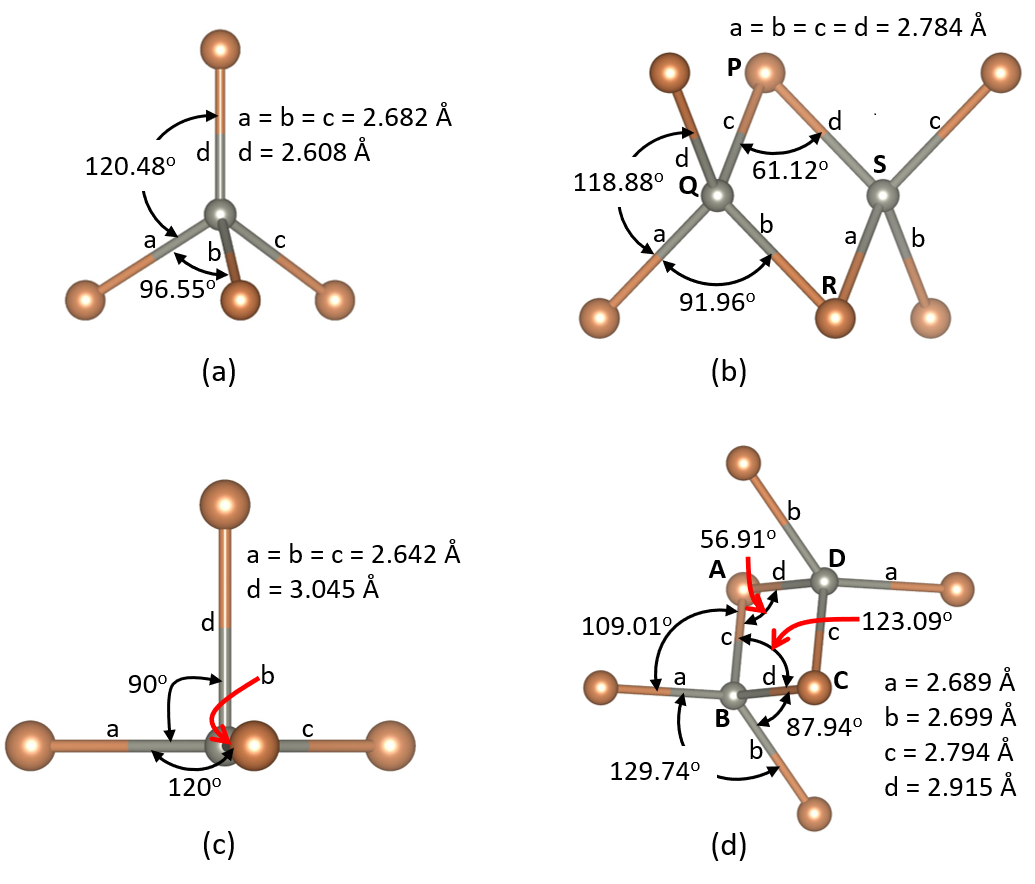}
\centering
\caption{Structural fragments showing tetrahedral configuration in (a) $w$-ZnSb, (b) $t$-ZnSb, (c) $h$-ZnSb, and (d) $o$-ZnSb).}\label{Fig:bond-length}
\end{figure*}
\begin{figure*}[h]
	\includegraphics[width=1\linewidth]{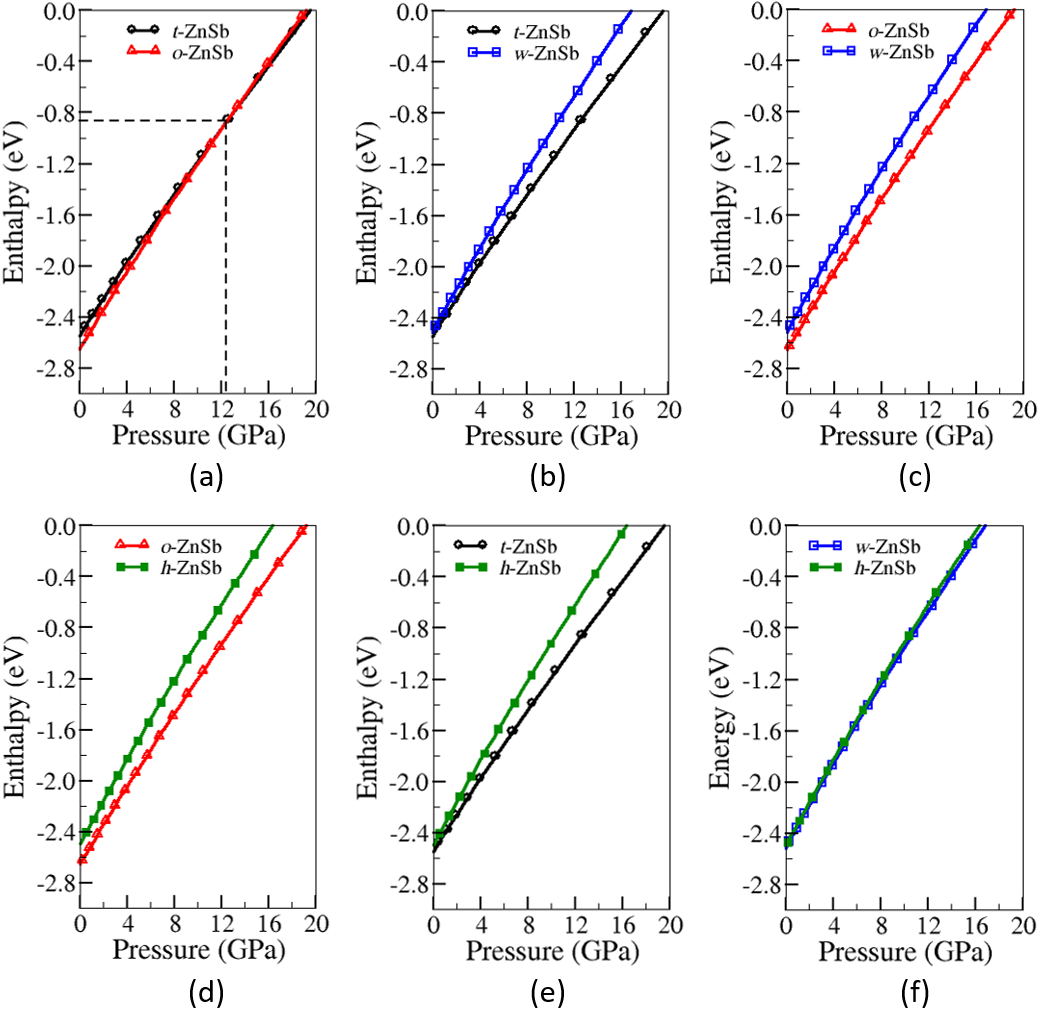}
	\centering
	\caption{Variation of calculated enthalpy (eV) per atom as a function of pressure at 0 K for $w$-ZnSb, $t$-ZnSb, $h$-ZnSb, and $o$-ZnSb.}\label{Fig:phase}
\end{figure*}

The equilibrium unit cell obtained from full structural optimization was expanded and compressed uniformly around the equilibrium volume. We optimized these expanded or compressed structures separately at constant volume and calculated self consistent energy value for each structure to generate a set of energy-volume data. Then, we fitted the total energy per unit cell as a function of volume using the $3^{rd}$ order Birch-Murnaghan~\cite{birch1947finite,garai2008universal} equation of state using Eqn.1 to determine the equilibrium value of bulk modulus ($B_o$), and pressure derivative ($B_{o}^{'}$) of the material.

\begin{equation}
\resizebox{0.8\linewidth}{!}{$
	E(V)= E_{o} + \frac{9V_{o}B_{o}}{16}\left[\Bigg\{\left(\frac{V_{o}}{V}\right)^{\frac{2}{3}}-1\Bigg\}^{3}B_{o}^{'}+\Bigg\{\left(\frac{V_{o}}{V}\right)^{\frac{2}{3}}-1\Bigg\}^{2}\Bigg\{6-4\left(\frac{V_{o}}{V}\right)^{\frac{2}{3}}\Bigg\}\right]
	$}
\end{equation}
where $E$ is the total energy, and $E_{o}$ the total equilibrium energy per unit cell, $B_{o}$ the equilibrium bulk modulus, $V$ the unit cell volume, $V_{o}$ the equilibrium volume of the unit cell and $B'_{o}$ the first derivative of the bulk modulus with respect to $V = V_o$. Our calculated value of bulk modulus for $o$-ZnSb (3D) is 47.25 GPa which is in good agreement with the value obtained by Philippe Jund et al.~\cite{jund2012physical}. The reported experimental value of Bulk modulus for $o$-ZnSb at 300K is 54.75 GPa~\cite{balazyuk2008bulk}. We further studied the enthalpy (H) as a function of pressure (P) in different phases of ZnSb using the VASPKIT program~\cite{wang2019vaspkit}. The stability of the particular structure is decided by the minima of the Gibbs free energy. At 0 K the free energy becomes equal to the enthalpy (H) of the system as given by Eqn.2.
\begin{equation}
H = E_{t} + PV
\end{equation}
At a given pressure, a stable structure has the enthalpy equal to its lowest value and the transition pressure ($P_{t}$) is computed at which the enthalpies for the two phases are equal. We calculated the enthalpies of each phase of ZnSb as a function of pressure as shown in Fig.~\ref{Fig:phase}(a-f). The orthorhombic $o$-ZnSb undergoes phase transition at $P_{t}$ = 12.48 GPa/atom to tetragonal $t$-ZnSb as shown in Fig.~\ref{Fig:phase}(a). So, at elevated pressure above 12.48 GPa/atom, $o$-ZnSb structure exists in tetragonal phase. Except $o$-ZnSb and $t$-ZnSb, all other structures undergo phase transition at negative pressures as represented in Table.\ref{Table:phase-transition}. Because of the dearth of experimental and theoretical values of transition pressure of these structures in the literature, our results serve as a complement for future investigation.

\begin{table}
	\centering
	\caption{Estimation of phase transition pressure $P_{t}$ (GPa/atom) and transition enthalpy $H_{t}$ (eV/atom).\label{Table:phase-transition}}
		\begin{tabular}{l*{3}{c}}
		\hline
			  Phase              & $P_{t}$(GPa/atom) & $H_{t}$ (eV/atom)  \\
			\hline
			$o$-ZnSb $\rightarrow$ $t$-ZnSb & 12.48 & -0.87  \\	
			$t$-ZnSb $\rightarrow$ $w$-ZnSb & -1.38 & -2.75  \\
			$o$-ZnSb $\rightarrow$ $w$-ZnSb & -9.95 & -4.22  \\
		    $o$-ZnSb $\rightarrow$ $h$-ZnSb & -10.94 & -4.38  \\
			$t$-ZnSb $\rightarrow$ $h$-ZnSb & -2.15 & -2.86  \\
			$w$-ZnSb $\rightarrow$ $h$-ZnSb & -46.11 & -10.44  \\
			\hline
		\end{tabular}
\end{table}

\subsection{II. Calculation of formation energy ($\Delta$E)}
In this section we reported the calculated formation energy of four different phases of ZnSb using Eqn.~\ref{Eqn:E-form}.

\begin{equation}
\resizebox{0.6\linewidth}{!}{$
	\Delta E(ZnSb)=  E(ZnSb)-\frac{1}{N}\left[N_{Zn}E(Zn)+N_{Sb}E(Sb)\right]
	$}\label{Eqn:E-form}
\end{equation}
where E(ZnSb), E(Zn) and E(Sb) are the calculated equilibrium energies in eV/atom of the corresponding bulk phases of ZnSb compounds in which hexagonal structures Zn($3d^{10}4s^{2}$) and Sb($5s^{2}5p^{3}$) elements belong to the space group $P6_{3}/mmc(n^{o}194)$, and $R\bar{3}m(n^{o}166)$ respectively. $N_{Zn}$ is the number of Zinc atoms, $N_{Sb}$ the number of Sb atoms, and N the total number of atoms including both Zn and Sb in the unit cell.  The formation energy $o$-ZnSb($pbca$) is found to be -0.032 eV/atom which is in good agreement with the previous calculated values~\cite{jund2012physical} but still overestimating the experimental values of -0.07 to -0.09 eV/atom~\cite{arushanov1986crystal}. Our results suggests that $t$-ZnSb is the most stable structure next to existing stable $o$-ZnSb. The computed formation energy for all the phases using DFT-GGA are tabulated in Table.~\ref{Table:E-form}. 

\subsection{III. Calculation of Exfoliation energy ($\rm {E_{exf}}$)}

\begin{figure}[h]
	\includegraphics[width=0.6\linewidth]{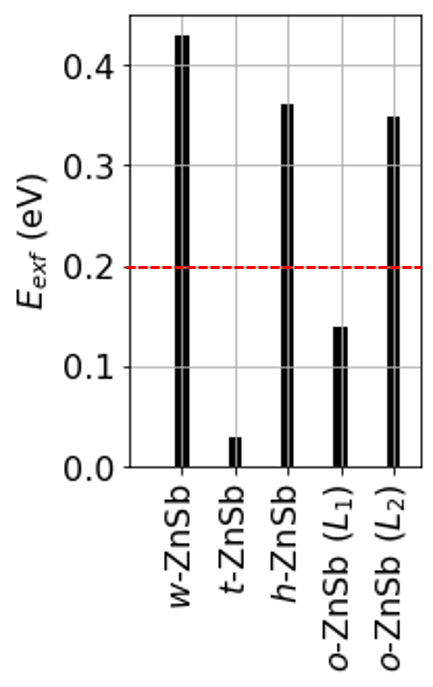}
	\centering
	\caption{The bar diagram showing the computed value of exfoliation energies in eV/atom for all 2D layered ZnSb structures. The red dashed line represents the exfoliation energy criterion.}\label{Fig:exf}
\end{figure}

The exfoliation energy to check the feasibility of the extraction of the 2D-monolayer was computed for all the bulk phases of ZnSb using Eqn.~\ref{Eqn:exf}following the procedures in the paper ~\cite{jung2018rigorous} :
\begin{equation}
	E_{exf} = \frac{E_{ML}}{N_{ML}}-\frac{E_{Bulk}}{N_{Bulk}} \label{Eqn:exf}
\end{equation}
where, $E_{ML}$, $N_{ML}$ are the total energy of the isolated primitive monolayer in vacuum and the number of atoms in the monolayer and $E_{Bulk}$, $N_{Bulk}$ are the total energy and the number of atoms in the bulk phases of the 2D layered ZnSb in their unit cell respectively. Fig~\ref{Fig:exf} represents the bar diagram for the calculated value of exfoliation energies in all the ZnSb structures. Because of anisotropic layer pattern in $o$-ZnSb, we attempted to imagine the monolayer of $o$-ZnSb breaking along the two different planes $L_{1}$ and $L_{2}$ as shown in Fig.~\ref{Fig:3D-bulk}(B). It has been observed that breaking $o$-ZnSb along $L_{1}$ plane is energetically favorable. The predicted energy difference while breaking through $L_{1}$ and $L_{2}$ is found to be $\Delta E$= E{($L_{1}$)}-E{($L_{2}$)} = -1.66 eV/unit cell, where E($L_{1}$) and E($L_{2}$) are the energy required to extract the monolayer of $o$-ZnSb breaking through $L_{1}$ and $L_{2}$ planes respectively. The obtained value of $E_{exf}$ suggests that the ML of $o$-ZnSb and $t$-ZnSb can be exfoliated since it falls under the conventional computational criterion to predict that $E_{exf}$ for a feasible 2D materials should be less than 0.2 eV/atom~\cite{choudhary2017high}. In addition, the monolayer of $o$-ZnSb is energetically feasible while breaking along the plane $L_{1}$. 
\begin{table}
	\centering
	\caption{Calculated Formation energy $\Delta$E in eV/atom.\label{Table:E-form}}
		\begin{tabular}{l*{2}{c}}
		\hline
			Compound                   & $\Delta$E  \\
			\hline
			$w$-ZnSb & 0.110 \\
			\hline
			$t$-ZnSb                     &  0.078 \\
			\hline
			$h$-ZnSb                     & 0.128 \\
			\hline
			$o$-ZnSb & -0.032\\
			\hline
		\end{tabular}
\end{table}

\subsection{IV. Electronic properties}
\begin{figure*}[h]
	\includegraphics[width=1\linewidth]{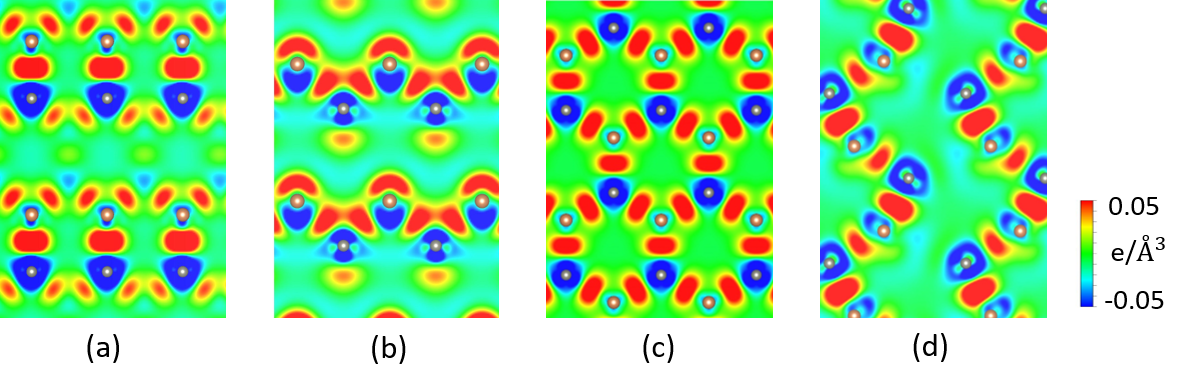}
	\centering
	\caption{2D contour plot of charge density difference in (a) $w$-ZnSb, (b) $t$-ZnSb, (c) $h$-ZnSb, and (d) $o$-ZnSb along (001) plane. The charge density difference value ranges from -0.05 (blue) to +0.05 (red) in the unit of $e/{\mbox{\normalfont\AA}}^{3}$.}\label{Fig:cd}
\end{figure*}

\begin{figure*}[h]
	\includegraphics[width=1\linewidth]{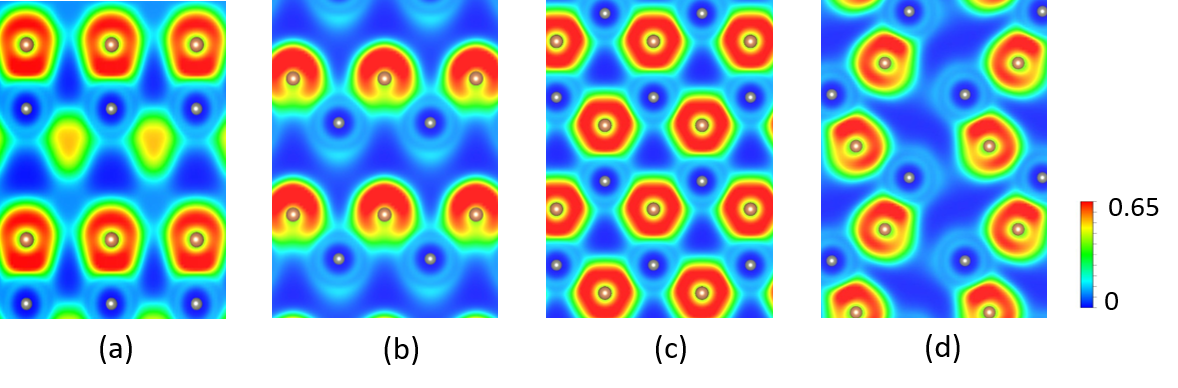}
	\centering
	\caption{2D contour plot of ELF in (a) $w$-ZnSb, (b) $t$-ZnSb, (c) $h$-ZnSb, and (d) $o$-ZnSb along (001) plane. The ELF value ranges from 0.00 (blue) to 0.65 (red).}\label{Fig:elf}
\end{figure*}

In this section, we reported our calculations on charge density difference($\Delta$ $\rho$), electron localization function (ELF), Bader charge analysis, total (T) and partial (P) density of states (DOS) for the different phases of ZnSb to elucidate their electronic properties. This subsection also includes the TDOS and PDOS as well as the electronic band structures of the energetically favored monolayer of $o$-ZnSb and $t$-ZnSb. The 2D plane for the contour plot of charge density difference and ELF is chosen along (001). The color scale on the plots follows that the low values are assigned dark blue and the red for the highest value. To discern a detail descriptions of the electron distribution around the atoms, and the nature of chemical bonding in them, the charge density difference and ELF are plotted for each structure as shown in Fig.~\ref{Fig:cd} and Fig.~\ref{Fig:elf}. The difference in charge density is calculated by subtracting charge densities of free Zn and Sb atoms from the total charge density of ZnSb using the relation $\Delta$ $\rho$ = $\rho_{ZnSb}$ - $\rho_{Zn}$ - $\rho_{Sb}$. Necessarily, we computed the value of $\rho_{Zn}$, and $\rho_{Sb}$ considering a fixed geometry or lattice parameters on which we calculated $\rho_{ZnSb}$. The value of $\Delta$ $\rho$ suggests that the charge density is localised in the midway of in-plane Zn-Sb bond indicating the strong covalent bonding between them. Similarly, ELF by definition refers to the probability of finding an electron in the neighborhood of another electron with the same spin. It takes the value ranging from 0.0 to 1.0 and 0.5 for the homogeneous electron gas. Thus, ELF of 1.0 corresponds to perfect localization and ELF of 0.5 to perfect delocalization. It has been a powerful tool since it was introduced in 1990 by Becke and Edgecombe~\cite{becke1990simple} to analyse the localization of parallel spin electrons in atoms, molecules, and solids providing the quantitative description of Pauli’s Exclusion Principle. In order to elucidate the region of localized electrons an ELF value of 0.65 is chosen in the present calculations. The ELF basins, the regions where the probability of finding a pair of electrons is maximum~\cite{savin2005significance}, are localized around the Sb atoms in all the structures indicating some degree of ionic nature in Zn-Sb bond and the strong covalency in Sb-Sb bond. This confirms that the electronegative Sb atom satisfies the octet configuration via the formation of lone pair localization with the electropositive Zn atom. 

The quantitative description of the type of bonding was performed by calculating the ionicity or the charge transfer in the compound using the Bader charges analysis. For this, the total charge density, including core charges was calculated. Atomic or Bader regions are defined as the 2D surfaces through which the gradient of the charge density has zero flux~\cite{benson2011electronic,tang2009grid}. The total charge on an atom was estimated by integrating the charge density within a region associated to a nucleus.The mesh for the augmentation charges was tested starting from the mesh size of the relaxation calculation~\cite{niedziolka2014theoretical}, and increasing it step-wise by 50\% up to 300\%. A grid size of 180$\times$180$\times$240 (200\%), 210$\times$210$\times$336 (250\%), 252$\times$252$\times$336 (250\%), and 288$\times$360$\times$384 (200\%) for $w$, $t$, $h$, and $o$-ZnSb respectively were enough to securely converge the Bader charge in the atoms~\cite{yu2011accurate}. 
we employed a weight method which offers the quadratic convergence as a function of grid density to perform the Bader charge partitioning for the electronic charge density as proposed by Min Yu, and Dallas R. Trinkle~\cite{yu2011accurate}. The weight method to calculate Bader charge is the efficient and accurate method compared to other grid based algorithms. Also, the calculated value of charge transfer between Zn and Sb in each of the structure agrees with the theoretical difference in Pauling electronegativity between Zn (1.65) and Sb (1.96)~\cite{nikitin2016thermoelectrics}. The value of charge transfer indicates that the $t$-ZnSb and $o$-ZnSb exhibits same degree of covalency whereas the $h$-ZnSb represents high degree of covalency as shown in Table.~\ref{Table:charge-transfer}. This is attributed to the change in $sp^{3}$ hybrid orbital in 3D-ZnSb to the $sp^{2}$ hybrid orbitals in honey comb structure of $h$-ZSb. 
\begin{table}
	\centering
	\caption{Bader charge analysis.\label{Table:charge-transfer}}
		\begin{tabular}{l*{2}{c}}
		\hline
			 Compound    & \multicolumn{2}{c}{\underline{Charge transfer (charge on atom)}} \\
			         & Zn & Sb \\ 
			 \hline
			$w$-ZnSb                     & +0.36   & -0.36 \\
			$t$-ZnSb                     & +0.34   & -0.34 \\
			$h$-ZnSb                     & +0.26   & -0.26 \\
			$o$-ZnSb                     & +0.34   & -0.34 \\
			\hline
		\end{tabular}
\end{table}

\begin{figure*}[h]
	\includegraphics[width=1\linewidth]{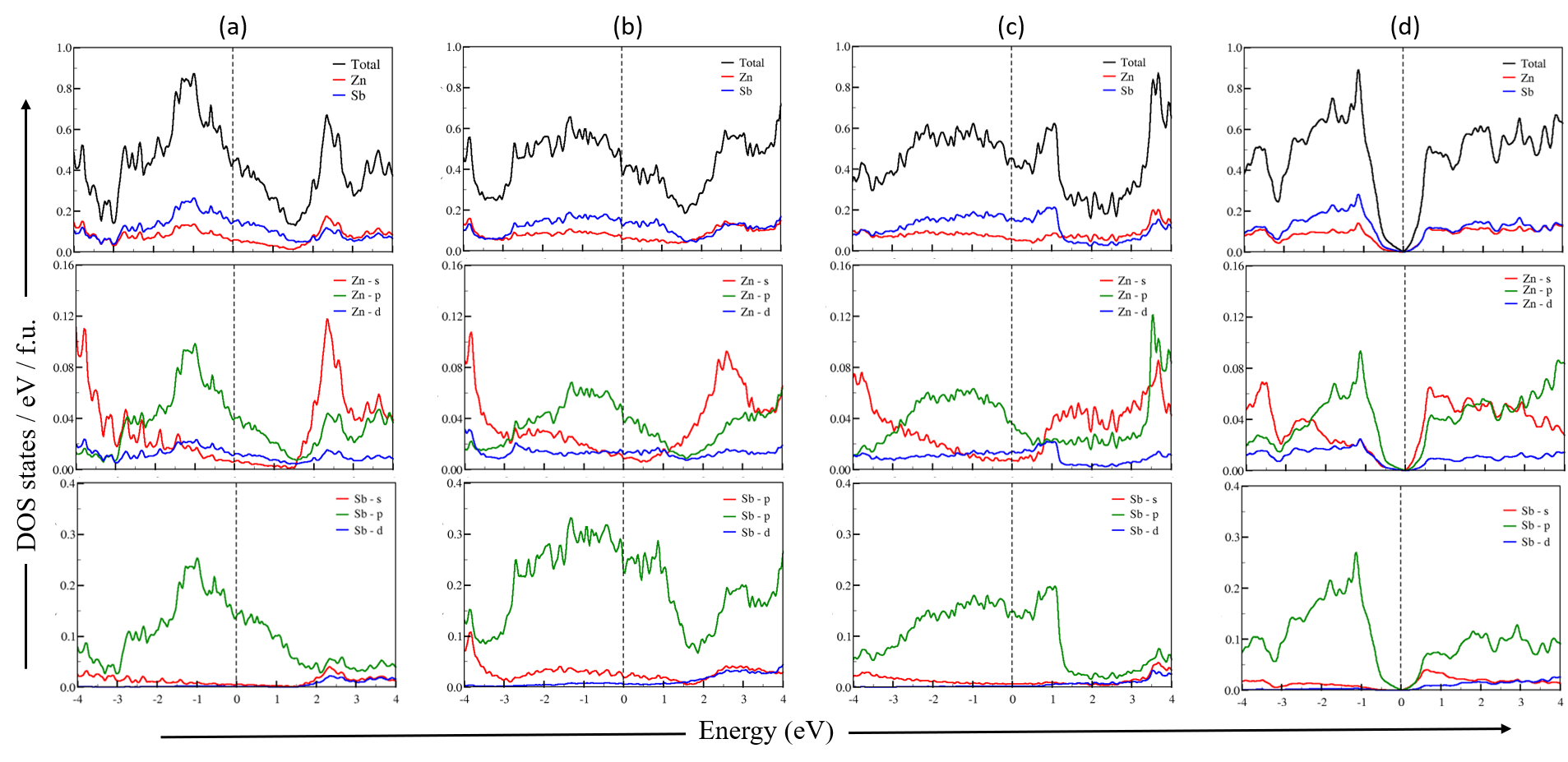}
	\centering
	\caption{Total DOS and orbital projected DOS of (a) $w$-ZnSb, (b) $t$-ZnSb, (c) $h$-ZnSb, and (d) $o$-ZnSb respectively calculated at T = 0 K. In all the plots, zero of the energy axis represents the Fermi level.}\label{Fig:dos}
\end{figure*}

\begin{figure*}[h]
	\includegraphics[width=1\linewidth]{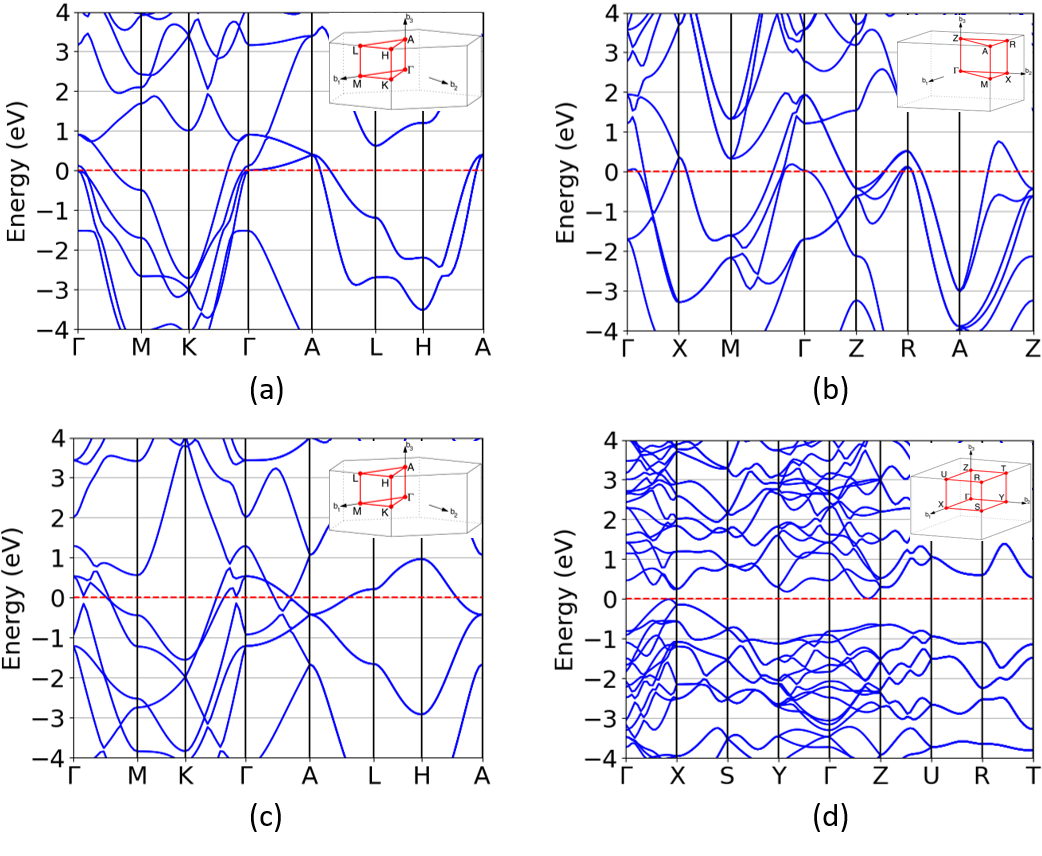}
	\centering
	\caption{Electronic band structure plots of (a) $w$-ZnSb, (b) $t$-ZnSb, (c) $h$-ZnSb, and (d) $o$-ZnSb at T = 0 K. Zero of energy axis represents the Fermi level. The inset diagram in each plot represents the first Brillouin zone showing high symmetry kpoints.}\label{Fig:band}
\end{figure*}

In order to get more insight on the electronic properties of ZnSb, we plotted the TDOS and PDOS of the ZnSb structures. The atom projected DOS reveals the fact that Sb-atom has major contribution near the Fermi level in all the ZnSb structures. In orbital projected  DOS of all the 2D-ZnSb, there is dominant contribution of Sb($p$) and Zn($p$) near the Fermi level. In case of $o$-ZnSb, Sb($p$) and Zn($p$) make the major contribution in the valence band edge. Similarly, in the conduction band edge, Sb($p$), Sb($s$) and Zn($p$), Zn($s$) make the dominant contribution. It is noteworthy that there is distinct $p-d$ hybridization between Sb($5p$) and Zn($3d$) in the studied range of -4.0 eV to +4.0 eV which is the cause of strong covalent bonding between Zn and Sb. It can be confirmed from the DOS and band structure plots that $w$, $t$, and $h$-ZnSb structures are metallic whereas, $o$-ZnSb is a semiconductor with an indirect, narrow band gap value of 0.03 eV as shown in Fig.~\ref{Fig:dos} and Fig.~\ref{Fig:band}. Here, the conduction band maximum (CBM) and valence band minimum (VBM) lies along $\Gamma$-Z and $\Gamma$-X directions of high symmetry k-path respectively characterized by multi-valley features in conduction band. The reported value of experimental band gap for $o$-ZnSb are 0.50, 0.59, and 0.69 eV at temperature 300 K, 77 K, and 4.2 K~\cite{komiya1964optical}.

Since, the monolayers of $t$-ZnSb and $o$-ZnSb are feasible based on the exfoliation energy criterion, we further studied the structural geometry and electronic properties of these 2D monolayers via DOS and band structures. Our optimized lattice parameters for ML of $t$-ZnSb (2D) is 3.812 \AA. The absolute change in lattice parameter ($a$) from its 2D layered bulk structure is $\delta a$ = 0.188 \AA. The number of Zn and Sb atoms in primitive ML of $t$-ZnSb is same as that of its 2D structure. The Zn-Sb bond length in 2D ML structure of $t$-ZnSb is 2.770 \AA~ which is slightly shorter than 2.784 \AA~ of its corresponding bulk structure. Similarly, we observed the significant change in lattice parameters in the 2D-ML structure of $o$-ZnSb from its bulk counterpart while breaking along energetically favorable plane $L_{1}$ after the geometrical relaxation. The optimized lattice parameters are $a$ = 4.544 \AA, $b$ = 8.812~\AA, and $\alpha$ = $90.39^{o}$, $\beta$ = $\gamma$ = $90^{o}$. Here, the absolute change in lattice parameter in 2D-ML structure from 3D-bulk structure along $a$ direction is $\delta$$a$ = 1.74 \AA~ and along $b$ direction is $\delta$$b$ = 0.99 \AA. The 2D-ML of $o$-ZnSb is characterized by 4 Zn and 4 Sb atoms constituting a chain of $Zn_{2}Sb_{2}$ parallelograms connected through the Sb-Sb dimers along $ab$ plane as shown in Fig.~\ref{Fig:3D-bulk}(B). The crystal symmetry of 2D-monolayer of $t$-ZnSb retains same as that of its bulk structure $i.e.$ $p4/nmm(n^{o}129)$; however, the crystal symmetry of 2D-monolayer of $o$-ZnSb is changed to $p2_{1}/c(n^{o}14)$. The total DOS and electronic band structures of 2D-monolayer of $t$-ZnSb and $o$-ZnSb reveals their metallic and semiconducting behavior respectively. Interestingly, the nature of band gap changes from indirect to direct while going from 3D bulk to 2D monolayer in $o$-ZnSb. Here, both the CBM and VBM lies along $\Gamma$-Z direction opening sizable and direct band gap of $\Delta E$ = 0.94 eV using GGA as shown in Fig.~\ref{Fig:band-mono}.        
 
  
\subsection{V. Phonon dispersion}
In order to study the lattice dynamics of ZnSb structures, we calculated the phonon dispersion relations using the phonopy code based on density functional perturbation theory (DFPT)~\cite{phonopy} along various high symmetry directions for all the bulk phases of ZnSb and their 2D-monolayer structures. Among the bulk phases, $t$-ZnSb has been found to be dynamically stable next to the existing stable phase of $o$-ZnSb as manifested by the absence of imaginary mode of vibration in phonon dispersion curves throughout the Brillouin zone. Similarly, among the 2D-monolayer cases, the monolayer of $o$-ZnSb is found to be dynamically stable as shown in Fig.~\ref{Fig:phonon-bulk} and \ref{Fig:phonon-mono} respectively.
\subsection{VI. Work function ($W_{f}$)}
In this section, we reported the work function of single layer thin slab of $o$-ZnSb and $t$-ZnSb. By definition, work function is the minimum energy required to remove an electron from the surface of the material as given by Koopmans' theorem in Eqn.5\cite{gu2015first,perdew1979accurate}, 
\begin{equation}
W_{f} = V_{vac}-E_{f}
\end{equation} 
where the potential in the vacuum region $V_{vac}$ and Fermi energy $E_{f}$ are derived from the same self consistent calculation. The potentials discussed in this calculation refers to the electrostatic part of the Hartree potential. The obtained value of work functions for these slab models using Eqn.5 are 4.04 eV and 4.61 eV for the semiconducting surface of $o$-ZnSb (3D) and metallic surface of $t$-ZnSb (2D) respectively as shown in Fig.~\ref{Fig:wf}. These monolayers can be utilized to make hetero-structures for electronic device applications. The metallic $t$-ZnSb can be used as contact layers in these devices. 
\begin{figure*}[h]
\includegraphics[width=1\linewidth]{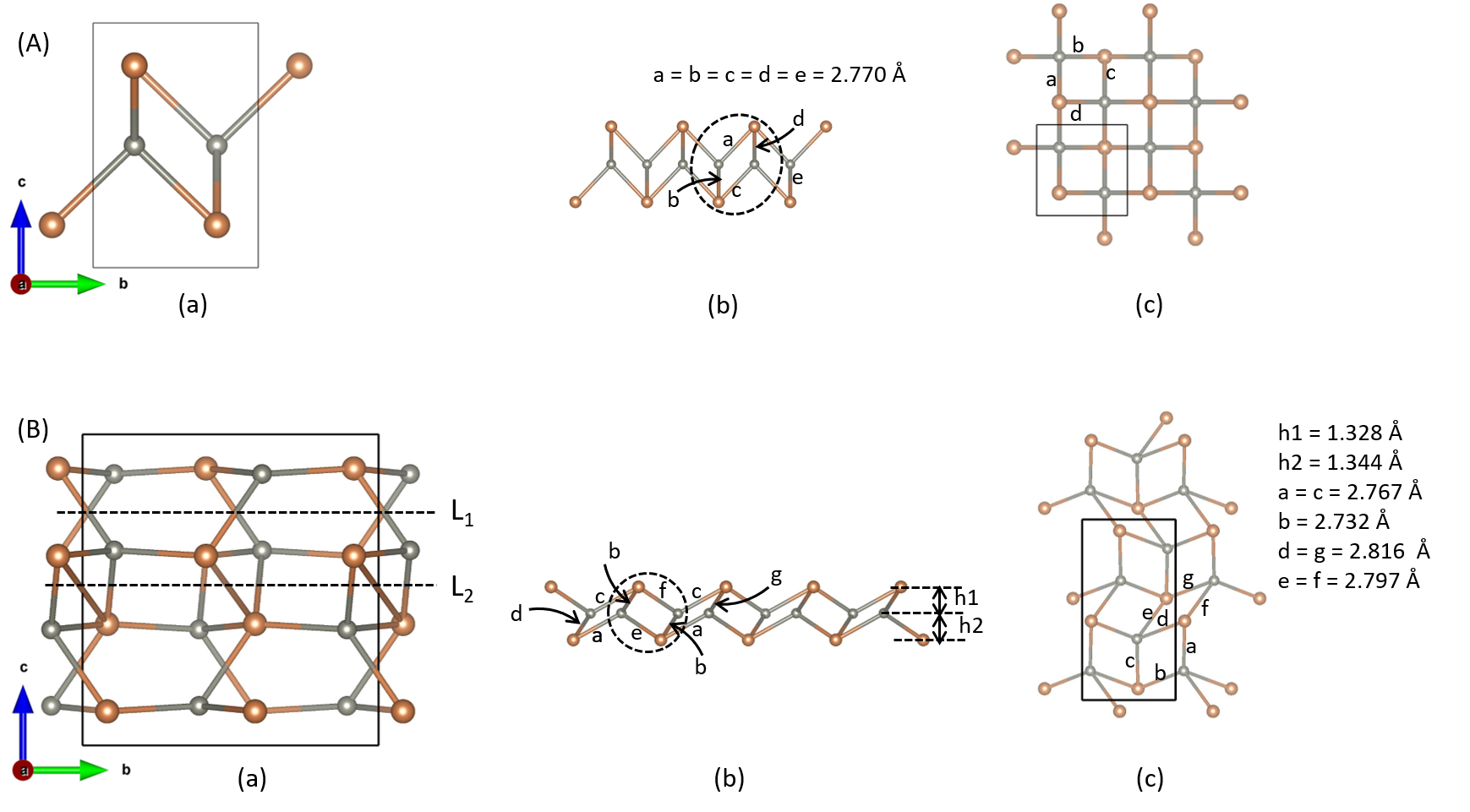}
\centering
\caption{(A) Schematic diagram of (a) bulk structure, (b) side view along (100) of 2D-monolayer, and (c) top view along (001) plane of 2D-monolayer structure in $t$-ZnSb. (B) Schematic diagram of (a) bulk structure showing possible breaking planes of $L_{1}$ and $L_{2}$, (b) side view along (100) of 2D-monolayer, and (c) top view  along (001) plane in $o$-ZnSb monolayer breaking through energetically easy plane $L_{1}$. The dotted circular ring represents $Zn_2Sb_2$ unit forming the rhomboid structure.}\label{Fig:3D-bulk}
\end{figure*}

\begin{figure*}[h]
	\includegraphics[width=1\linewidth]{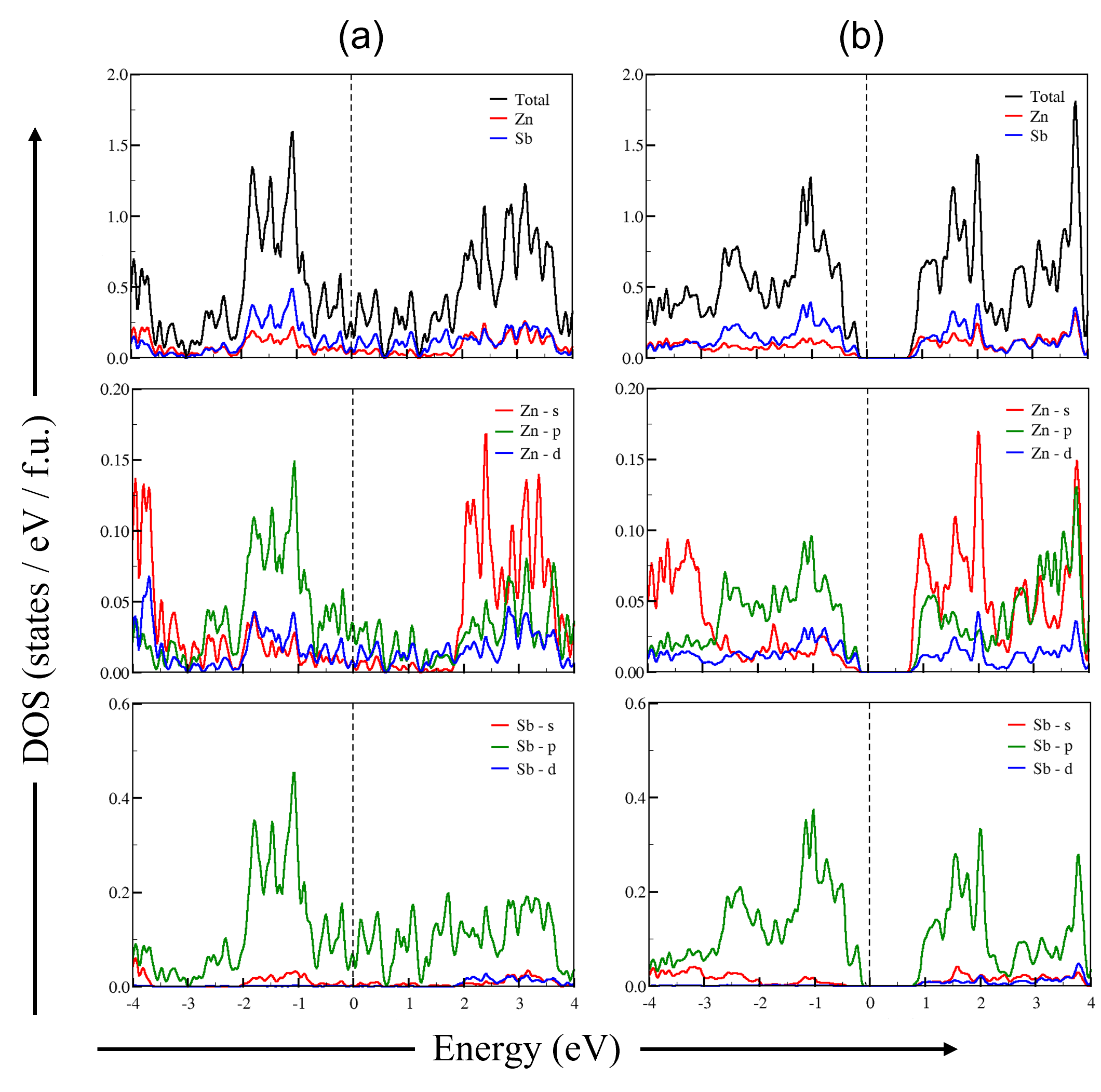}
	\centering
	\caption{Total DOS and orbital projected DOS of 2D-monolayer of (a) $t$-ZnSb, and (b) $o$-ZnSb calculated at T = 0 K. In all the plots, zero of the energy axis represents the Fermi level.}\label{Fig:dos-mono}
\end{figure*}
\begin{figure}[h]
	\includegraphics[width=1\linewidth]{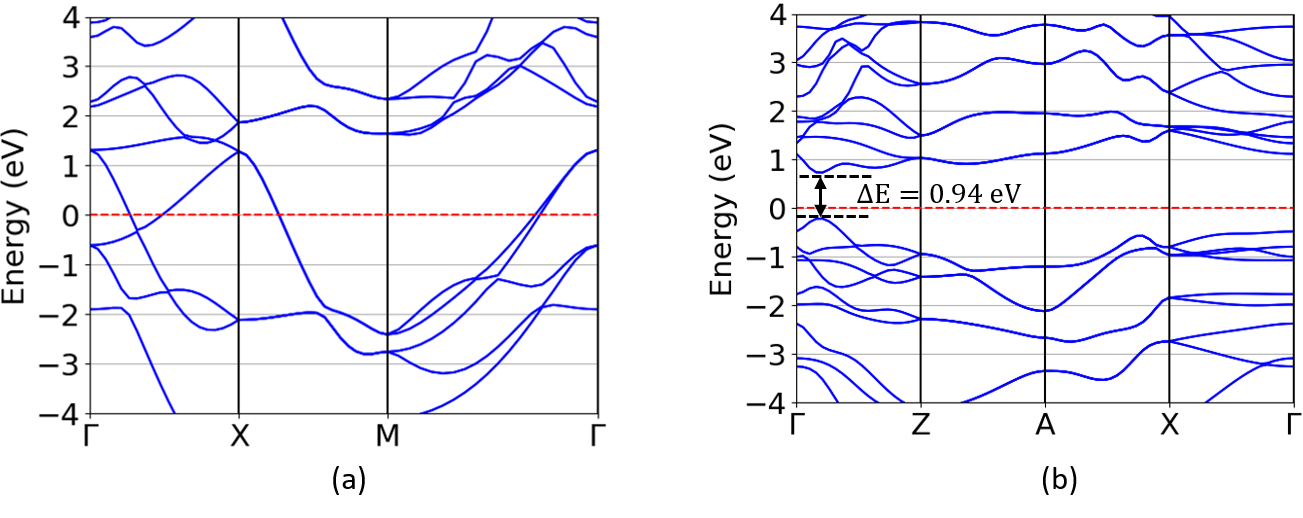}
	\centering
	\caption{Electronic band structures of 2D monolayer of (a) $t$-ZnSb, and (b) $o$-ZnSb calculated at T = 0 K. Zero of the energy axis represents the Fermi level.}\label{Fig:band-mono}
\end{figure}
\begin{figure*}[h]
	\includegraphics[width=1\linewidth]{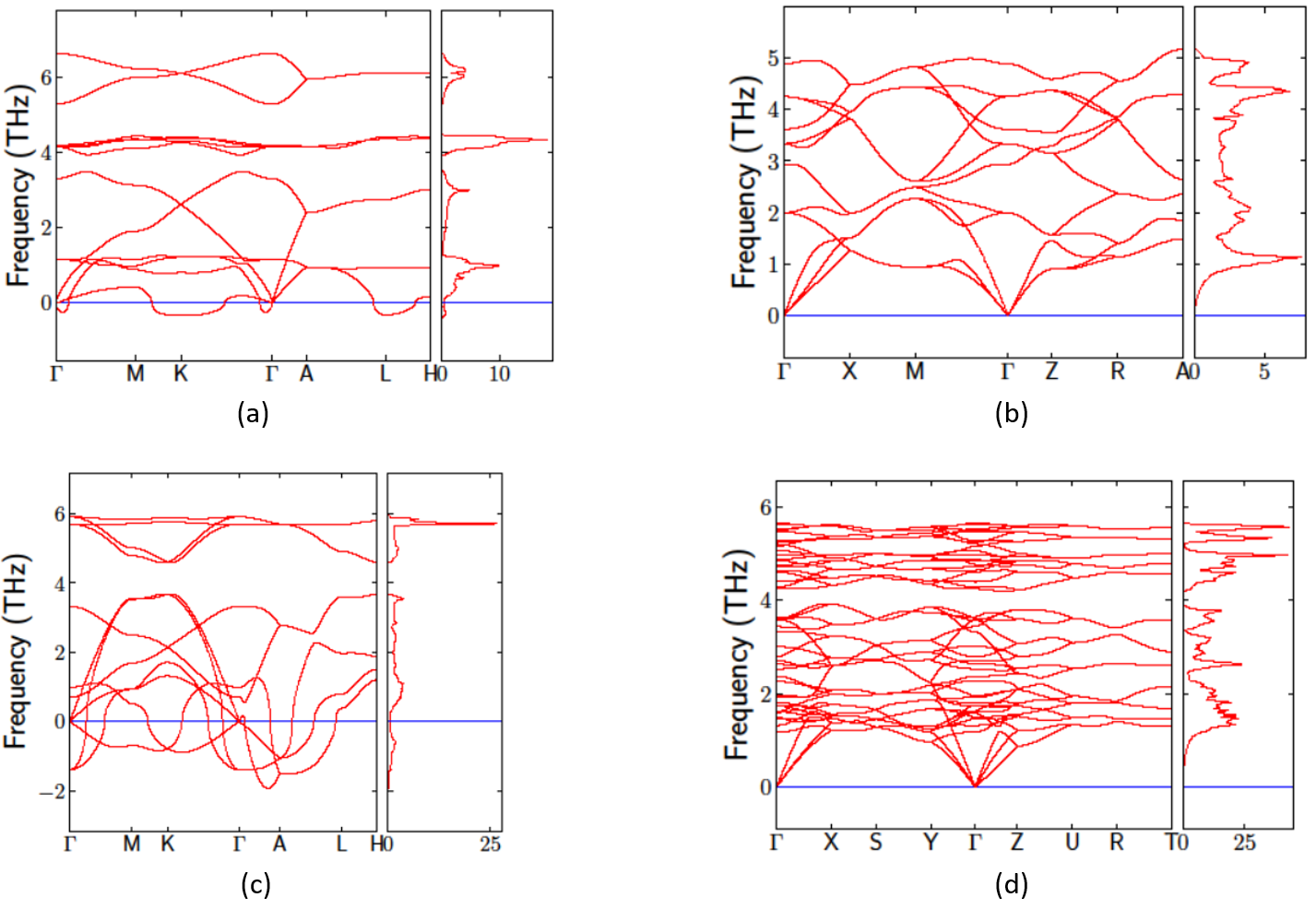}
	\centering
	\caption{Phonon dispersion curves for bulk structures of (a) $w$-ZnSb, (b) $t$-ZnSb, (c) $h$-ZnSb, and (d) $o$-ZnSb.}\label{Fig:phonon-bulk}
\end{figure*}

\begin{figure*}[h]
	\includegraphics[width=1\linewidth]{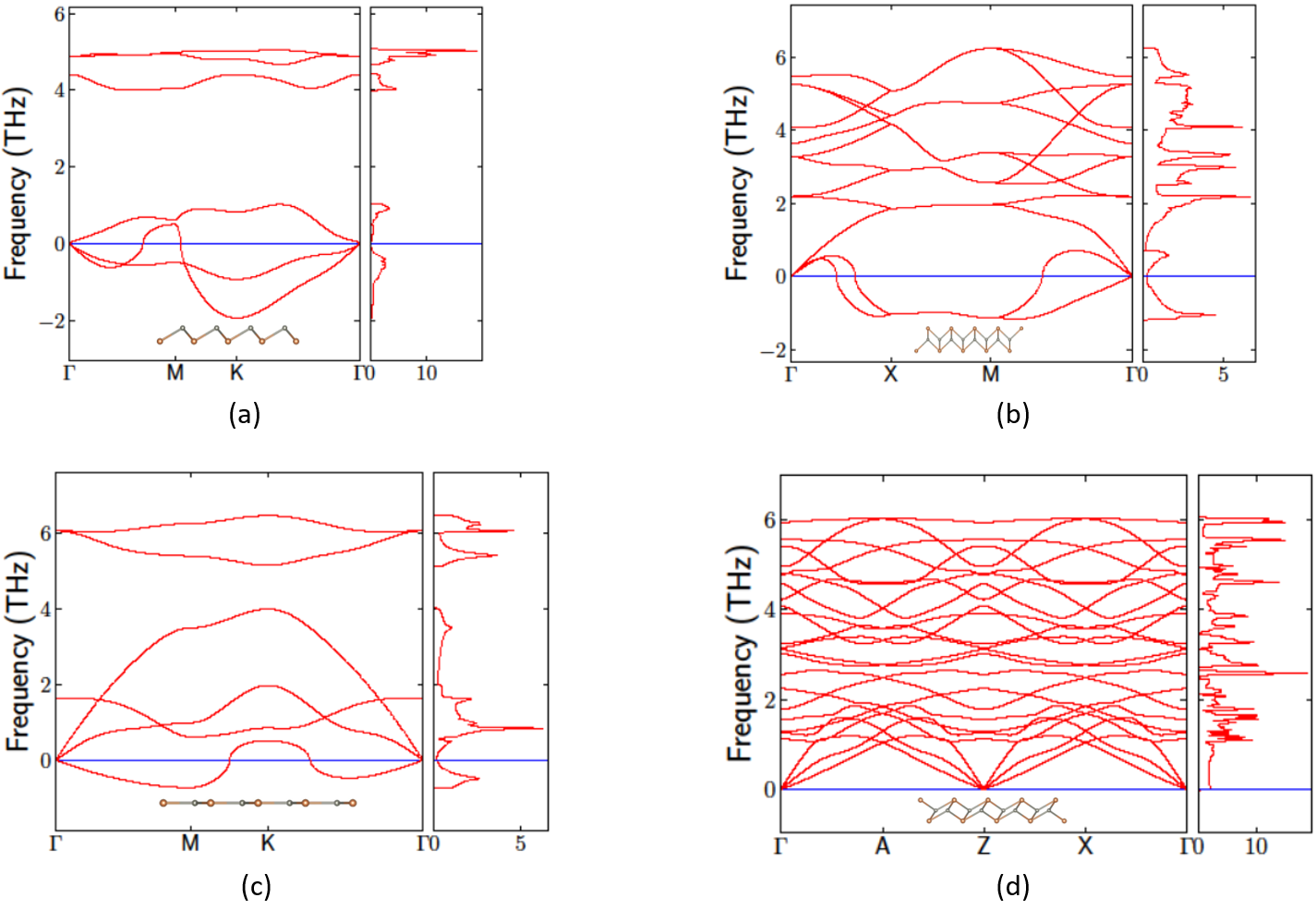}
	\centering
	\caption{Phonon dispersion curves for 2D-monolayer structures of (a) $w$-ZnSb, (b) $t$-ZnSb, (c) $h$-ZnSb, and (d) $o$-ZnSb.}\label{Fig:phonon-mono}
\end{figure*}
\begin{figure*}[h]
	\includegraphics[width=1\linewidth]{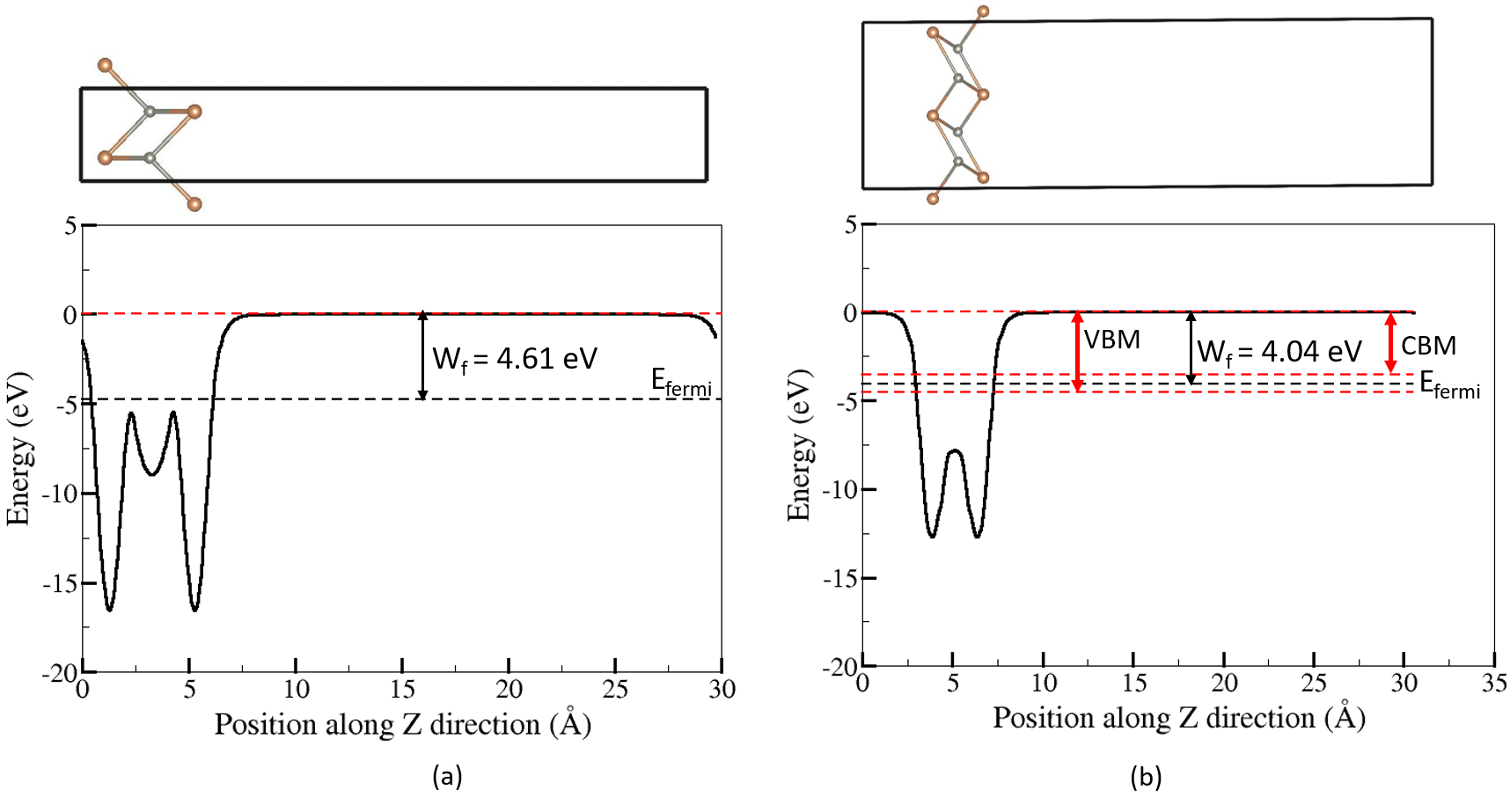}
	\centering
	\caption{Electrostatic potential energy variation along $z$-direction for a single layer slab of (a) $o$-ZnSb and (b) $t$-ZnSb. Here, the black dotted line represents the Fermi level. The potential energy of the vacuum level is shifted at zero.}\label{Fig:wf}
\end{figure*}

\section{Conclusion}
In summary, we have studied the structural, electronic, and phase stability of 2D layered structures that belong to wurzite-$w$, tetragonal-$t$, and hexagonal-$h$ phases, and 3D orthorhombic-$o$ phases of ZnSb using the first principle calculations. The 2D layered structures of ZnSb were modeled from the de-intercalation of alkali metals (A) in AZnSb (A= Li, Na, and K) and relaxed fully using DFT-GGA to get their stable configurations. In this paper, we predicted the novel 2D layered structure of ZnSb, stabilized in tetragonal symmetry with the space group $P4/nmm (n^o186)$. We reported the pressure induced phase transition between ZnSb (orthorhombic) to ZnSb (tetragonal) at 12.48 GPa/atom. Our calculations show that at elevated pressure above 12.48 GPa/atom, the orthorhombic phase is stable in tetragonal structure. There is also small average value of charge transfer between Zn and Sb ranging from $\pm$0.26 ($h$-ZnSb) to $\pm$ 0.36 ($w$-ZnSb) indicating some degree of ionic bonding between them. 
In addition, we predicted the feasibility of novel 2D-monolayer structures of $t$-ZnSb and $o$-ZnSb based on the exfoliation energy criterion. The DOS and band structures plot reveal that the 2D-ML structures of $t$-ZnSb and $o$-ZnSb exhibit metallic and semiconducting behavior respectively. The nature of band gap in $o$-ZnSb changes from indirect (narrow) to direct (sizable) while going from bulk 3D to 2D monolayer which is the surprising result of our calculations. It is noteworthy that the novel bulk phase of $t$-ZnSb and 2D-monolayer structure of $o$-ZnSb are dynamically stable due to the absence of imaginary frequencies in phonon dispersion bands; however, the 2D-monolayer structure of $t$-ZnSb is not dynamically stable even though it satisfies the exfoliation energy criterion at ambient pressure and temperature of P = 0 GPa and T = 0 K respectively. Finally, we reported a low work function value of 4.04 eV and 4.61 eV for the single 2D slab of $o$-ZnSb and $t$-ZnSb respectively.The predicted low work function value and direct semiconducting behavior of 2D-monolayer structure of $o$-ZnSb signifies its promising applications in modern electronic devices.  

\section{Acknowledgements}
This work was supported by a National Research Foundation of Korea (NRF) grant funded by the Korean government (Ministry of Science, ICT \& Future Planning) (No.2015M3D1A1070639) and in part by the Center for Computational Sciences (CCS) at Mississippi State University.

\section{Conflict of interests}
The authors declare no conflict of interests.

\bibliography{references}

\end{document}